\documentclass[preprint,epsfig]{revtex4}
\usepackage{graphicx}
\begin{document}
\preprint{}
\title{SPINODAL INSTABILITIES IN NUCLEAR MATTER IN A STOCHASTIC RELATIVISTIC MEAN-FIELD APPROACH}

\author{S. Ayik$^{1}$}\email{ayik@tntech.edu}
\author{O. Yilmaz$^{2}$}
\author{N. Er$^{2}$}
\author{A. Gokalp$^{2}$}
\author{P. Ring$^{3}$}
\affiliation{$^{1}$Physics Department, Tennessee Technological University, Cookeville, TN 38505, USA \\
$^{2}$Physics Department, Middle East Technical University, 06531
Ankara, Turkey \\
$^{3}$Physics Department, TU Munich, D-85748 Garching, Germany}
\date{\today}

\begin{abstract}
Spinodal instabilities and early growth of baryon density
fluctuations in symmetric nuclear matter are investigated in the
basis of stochastic extension of relativistic mean-field approach in
the semi-classical approximation. Calculations are compared with the
results of non-relativistic calculations based on Skyrme-type
effective interactions under similar conditions. A qualitative
difference appears in the unstable response of the system: the
system exhibits most unstable behavior at higher baryon densities
around $\rho_{b}=0.4 ~\rho_{0}$ in the relativistic approach while
most unstable behavior occurs at lower baryon densities around
$\rho_{b}=0.2 ~\rho_{0}$ in the non-relativistic calculations

\end{abstract}

\maketitle

\section{Introduction}

Spinodal instability provides a possible dynamical mechanism for
fragmentation of a hot piece of nuclear matter produced in heavy-ion
collisions. Small amplitude density fluctuations grow rapidly and
lead to break-up of the system into an ensemble of clusters
\cite{R1}. In coming years, experimental investigations of
multi-fragmentation reactions in neutron rich nuclear system will
provide further understanding of isospin dependence of nuclear
matter equation of state at low densities.  In theoretical side,
extensive investigations of spinodal instabilities have been carried
out in the basis of stochastic transport models
\cite{R2,R3,R4,R5,R6}. In particular, the recently proposed
stochastic mean-field approach provides a useful tool for a
description of dynamics of density fluctuations in the spinodal
region \cite{R7}. It has been demonstrated that the stochastic
mean-field approach incorporates the one-body dissipation and the
associated fluctuation mechanism in accordance with the
quantal-dissipation fluctuation relation. The approach gives rise to
the same result for dispersion of one-body observables that was
obtained in a variational approach in a previous work \cite{R8}.
Furthermore, in recent studies \cite{R9,R10} by projecting onto
macroscopic variables, we deduce transport coefficients for energy
dissipation and nucleon exchange in low-energy heavy-ion collisions,
which have the similar form with those familiar from the
phenomenological nucleon exchange model \cite{R11}. These
investigations provide a strong support for the fact that the
stochastic mean-field approach is a powerful tool for describing low
energy nuclear collisions and spinodal dynamics.

In a recent work, we studied the early development of spinodal
dynamics of nuclear matter in the basis of the stochastic mean-field
approach by employing density-dependent Skyrme-type effective
interactions \cite{R12}.  In the present work, we carry out a
similar investigation of early development of density fluctuations
in spinodal region of nuclear matter by employing the stochastic
extension of the relativistic mean-field theory \cite{R13,R14}. It
has been shown in recent years that the nuclear many-body system is
in principal a relativistic system driven by dynamics of large
relativistic attractive scalar and repulsive vector fields. Both
fields are not much smaller than the nucleon mass and therefore the
average nuclear field should be described by Dirac equation.  For
large components of Dirac spinors, two fields nearly cancel each
other leading to relatively small attractive mean field. The small
components add up leading to a very large spin orbit term, which is
known since early days of nuclear physics. Relativistic models have
been used with great success to describe nuclear structure. In
recent years, the approach has also been applied for description of
nuclear dynamics extended in the framework of time-dependent
covariant density functional theory \cite{R15,R16}.   A number of
investigations have been carried out on spinodal instabilities in
nuclear matter employing relativistic mean-field approaches
\cite{R17,R18,R19}. In this work, we consider the stochastic
extension of the relativistic mean-field theory in the
semi-classical approximation. As illustrated in the non-relativistic
limit, stochastic extension of the mean-field theory provides a
powerful approach for investigating dynamics of density
fluctuations. Employing the stochastic extension of the relativistic
mean-field approach, we investigate not only spinodal instabilities
but also the early development of density fluctuations in symmetric
nuclear matter.

In Section 2, we briefly describe the stochastic extension of the
relativistic mean-field theory in the semi-classical approximation.
In Section 3, we calculate early growth of baryon density
fluctuations, growth rates and phase diagram of dominant modes in
symmetric nuclear matter. Conclusions are given in Section 4.

\section{STOCHASTIC RELATIVISTIC MEAN-FIELD THEORY}

The stochastic mean-field approach is based on a very appealing
stochastic model proposed for describing deep-inelastic heavy-ion
collisions and sub-barrier fusion \cite{R20,R21,R22}. In that model,
dynamics of relative motion is coupled to collective surface modes
of colliding ions and treated in a classical framework. The initial
quantum zero point and thermal fluctuations are incorporated into
the calculations in a stochastic manner by generating an ensemble of
events according to the initial distribution of collective modes. In
the mean-field evolution, coupling of relative motion with all other
collective and non-collective modes are automatically taken into
account. In the stochastic extension of the mean-field approach, the
zero point and thermal fluctuations of the initial state are taken
into account in a stochastic manner, in a similar manner presented
in refs. \cite{R20,R21,R22}. The initial fluctuations, which are
specified by a specific Gaussian random ensemble, are simulated by
considering evolution of an ensemble of single-particle density
matrices. It is possible to incorporate quantal and thermal
fluctuations of the initial state into  the relativistic mean-field
description in a similar manner.

In refs. \cite{R23,R24}, the authors derived a relativistic Vlasov
equation from the Walecka model in  the local density and the
semi-classical approximation. In the Walecka model, interaction
between nucleons are mediated by  a scalar meson  with mass $m_{s}$
and a vector meson with mass $m_\textrm{v}$, with respective fields
denoted as $\phi$  and $V_{\mu}$. Introducing phase space
distribution function $f(\vec{r},\vec{p},t)$ for the nucleons,
following relativistic Vlasov equation has been obtained ,
\begin{equation}\label{e1}
    \frac{\partial}{\partial t}f(\vec{r},\vec{p},t)+\vec{\textrm{v}}\cdot\vec{\nabla}_{r} f(\vec{r},\vec{p},t)-
    \vec{\nabla}_{r}
    h(\vec{r},\vec{p})\cdot\vec{\nabla}_{p}f(\vec{r},\vec{p},t)=0.
\end{equation}
where  $\vec{\textrm{v}}=\vec{p}^{*}/e^{*}$ and
$h=e^{*}+g_{\textrm{v}}V_{0}$. The coupling constants of the mesons
and the nucleon are denoted by $g_{s}$ and $g_{\textrm{v}}$, for the
scalar and the vector mesons, respectively. In these expressions,
$\vec{p}^{*}=\vec{p}-g_{\textrm{v}}\vec{V}$ and
$e^{*}=(\vec{p}^{*2}+M^{*2})^{1/2}$ with
$M^{*}=M-g_{\textrm{s}}\phi$. The nucleon mass is denoted by $M$. In
the mean-field approximation, the meson fields are treated as
classical fields and their evolutions are determined by the field
equations,
\begin{equation}\label{e2}
    \left[\frac{\partial^{2}}{\partial {t}^{2}}-\nabla^{2}+m_{\textrm{s}}^{2}\right]\phi(\vec{r},t)=g_{\textrm{s}}\rho_{\textrm{s}}(\vec{r},t)
\end{equation}
and
\begin{equation}\label{e3}
    \left[\frac{\partial^{2}}{\partial {t}^{2}}-\nabla^{2}+m_{\textrm{v}}^{2}\right]V_{\textrm{v}}(\vec{r},t)=g_{\textrm{v}}\rho_{\textrm{v}}(\vec{r},t).
\end{equation}
In these expressions, the baryon density
$\rho_{0}(\vec{r},t)=\rho_{\textrm{b}}(\vec{r},t)$, the scalar
density $\rho_{\textrm{s}}(\vec{r},t)$, and the current density
$\vec{\rho}_{\textrm{v}}(\vec{r},t)$ can be expressed in terms of
phase-space distribution function as follows,
\begin{eqnarray}\label{e4-5}
    \rho_{\textrm{b}}(\vec{r},t)=\gamma\int \frac{d^{3}p}{(2\pi)^{3}} f(\vec{r},\vec{p},t),\\
    \rho_{\textrm{s}}(\vec{r},t)=\gamma\int  \frac{d^{3}p}{(2\pi)^{3}}
    \frac{M^{*}}{e^{*}}f(\vec{r},\vec{p},t),
\end{eqnarray}
and
\begin{equation}\label{e6}
    \vec{\rho}_{\textrm{v}}(\vec{r},t)=\gamma\int  \frac{d^{3}p}{(2\pi)^{3}}
    \frac{\vec{p}^{*}}{e^{*}}f(\vec{r},\vec{p},t)
\end{equation}
where $\gamma=4$ is the spin-isospin degeneracy factor.  The
original Walecka model gives a nuclear compressibility that is much
larger than the one extracted from the giant monopole resonances in
nuclei. It also leads to an effective nucleon mass which is smaller
than the value determined from the analysis of nucleon-nucleon
scattering. In order to have a model which allows different values
of nuclear compressibility and the nucleon effective mass, it is
possible to improve the Walecka model by including the
self-interaction of the scalar mesons or by considering density
dependent coupling constants. However, in the present exploratory
work, we employ the original Walecka model without including the
self interaction of the scalar meson.

In the stochastic mean-field approach an ensemble
$\{f^{\lambda}(\vec{r}, \vec{p}, t)\}$ of the phase-space
distributions is generated in accordance  with the initial
fluctuations, where $\lambda$ indicates the event label.  In the
following for simplicity of notation, since equations of motions do
not change in the stochastic evolution, we do not use the event
label $\lambda$ for the phase-space distributions and also on the
other quantities. However it is understood that the phase-space
distribution, scalar meson and vector meson fields are fluctuating
quantities.  Each member of the ensemble of phase-space
distributions evolves by the same Vlasov \cite{R1} equation
according to its own self-consistent mean-field, but with different
initial conditions. The main assumption of the approach in the semi-
classical representation is the following: In each phase-space cell,
the initial phase-space distribution $f(\vec{r}, \vec{p}, 0)$ is a
Gaussian random number with its mean value determined by
$\overline{f(\vec{r}, \vec{p}, 0)}=f_{0}(\vec{r}, \vec{p})$, and its
second moment is determined by \cite{R7,R12}
\begin{equation}\label{e7}
    \overline{f(\vec{r}, \vec{p}, 0)f(\vec{r}', \vec{p}', 0)}=
    (2\pi)^3\delta(\vec{r}-\vec{r}')\delta(\vec{p}-\vec{p}')f_{0}(\vec{r},
    \vec{p})[1-f_{0}(\vec{r}, \vec{p})]
\end{equation}
where the overline represents the ensemble averaging and
$f_{0}(\vec{r}, \vec{p})$ denotes the average phase-space
distribution describing the initial state. In the special case of a
homogenous initial state, it is given by the Fermi-Dirac
distribution $f_{0}(p)=1/[\exp(e_{0}^{*}-\mu_{0}^{*})/T+1]$. In this
expression
$\mu_{0}^{*}=\mu_{0}-(g_{\textrm{v}}/m_{\textrm{v}})^{2}\rho_{\textrm{b}}^{0}$
where $\mu_{0}$ is the chemical potential and
$\rho_{\textrm{b}}^{0}$ is the baryon density in the homogenous
initial state.

In this work, we investigate the early growth of density
fluctuations in the spinodal region in symmetric nuclear matter. For
this purpose, it is sufficient to consider the linear response
treatment of dynamical evolution. The small amplitude fluctuations
of the phase-space distribution $\delta f(\vec{r},\vec{p},t)
=f(\vec{r},\vec{p},t)-f_{0}(\vec{p})$ around an equilibrium state
$f_{0}(\vec{p})$ are determined by the linearized Vlasov equation,
\begin{equation}\label{e8}
    \frac{\partial}{\partial t}\delta f(\vec{r},\vec{p},t)+\vec{\textrm{v}}_{0}\cdot\vec{\nabla}_{r}\delta f(\vec{r},\vec{p},t)-
    \vec{\nabla}_{r}\delta
    h(\vec{r},\vec{p},t)\cdot\vec{\nabla}_{p}f_{0}(p)=0.
\end{equation}
In these expression the local velocity is
$\vec{\textrm{v}}_{0}=\vec{p}/e^{*}_{0}$ with
$e^{*}_{0}=\sqrt{\vec{p}^{2}+M^{*2}_{0}}$,
$M_{0}^{*}=M-g_{\textrm{s}}\phi_{0}$, and small fluctuations of
mean-field Hamiltonian is given by,
\begin{eqnarray}\label{e9}
    \delta h(\vec{r},\vec{p},t) =-\frac{M_{0}^{*}}{e_{0}^{*}}g_{\textrm{s}}\delta \phi(\vec{r},t)
    +g_{\textrm{v}}\delta V_{0}(\vec{r},t)-\frac{g_{\textrm{v}}}{e_{0}^{*}}\vec{p}\cdot\delta \vec{V}(\vec{r},t)
\end{eqnarray}
The small fluctuations of the scalar and vector mesons are
determined by the linearized field equations,
\begin{equation}\label{e10}
    \left[\frac{\partial^{2}}{\partial {t}^{2}}-\nabla^{2}+m_{\textrm{s}}^{2}\right]\delta\phi(\vec{r},t)=g_{\textrm{s}}\delta\rho_{\textrm{s}}(\vec{r},t)
\end{equation}
and
\begin{equation}\label{e11}
    \left[\frac{\partial^{2}}{\partial {t}^{2}}-\nabla^{2}+m_{\textrm{v}}^{2}\right]\delta \vec{V}_{\textrm{v}}(\vec{r},t)=g_{\textrm{v}}\delta\vec{\rho}_{\textrm{v}}(\vec{r},t).
\end{equation}

\section{EARLY GROWTH OF DENSITY FLUCTUATIONS}

\subsection{Spinodal Instabilities}
In this section, we employ the stochastic relativistic mean-field
approach in small amplitude limit to investigate spinodal
instabilities in symmetric nuclear matter. We can obtain the
solution of linear response equations (\ref{e7})-(\ref{e11}) by
employing the standard method of one-sided Fourier transform in time
\cite{R25}. It is also convenient to introduce the Fourier transform
of the phase-space distribution in space,
\begin{eqnarray}\label{e12}
    \delta\tilde{f}(\vec{k},\vec{p},\omega)=\int_{0}^{\infty}dt e^{i\omega t}\int_{-\infty}^{\infty}d^{3}r
    e^{-i\vec{k}\cdot\vec{r}}f(\vec{r},\vec{p},t).
\end{eqnarray}
This leads to,
\begin{eqnarray}\label{e13}
    \delta\tilde{f}(\vec{k},\vec{p},\omega)=
    \left(\frac{M_{0}^{*}}{e_{0}^{*}}\tilde{g}_{\textrm{s}}^{2}\delta\tilde{\rho}_{\textrm{s}}(\vec{k},\omega)
    -\tilde{g}_{\textrm{v}}^{2}\delta\tilde{\rho}_{\textrm{b}}(\vec{k},\omega)+
    \tilde{g}_{\textrm{v}}^{2}\frac{\vec{p}}{e_{0}^{*}}\cdot\delta\tilde{\vec{\rho}}_{\textrm{v}}(\vec{k},\omega)\right)
    \frac{\vec{k}\cdot\vec{\nabla}_{p}f_{0}(p)}{\omega-\vec{\textrm{v}}_{0}\cdot\vec{k}}
    +i\frac{\delta\tilde{f}(\vec{k},\vec{p},0)}{\omega-\vec{\textrm{v}}_{0}\cdot\vec{k}}
    \nonumber \\
\end{eqnarray}
where $\delta\tilde{f}(\vec{k},\vec{p},0)$ denotes the Fourier
transform of the initial fluctuations, and we use the short hand
notation,
$\tilde{g}_{\textrm{s}}^{2}=g_{\textrm{s}}^{2}/(k^{2}+m_{\textrm{s}}^{2})$,
$\tilde{g}_{\textrm{v}}^{2}=g_{\textrm{v}}^{2}/(k^{2}+m_{\textrm{v}}^{2})$.
In this expression, the fluctuations of the meson fields are
expressed in terms of Fourier transforms of the scalar density
$\delta{\rho}_{s}(\vec{r}, t)$, the baryon density
$\delta{\rho}_{b}(\vec{r},t)$  and the current density
$\delta\vec{{\rho}}_{\textrm{v}}(\vec{r},t)$ fluctuations by
employing the field equations (\ref{e10})-(\ref{e11}). In Eq.
(\ref{e13}) only the initial fluctuations of the phase-space
distribution $\delta\tilde{f}(\vec{k},\vec{p},0)$ is kept, but the
initial fluctuations associated with the scalar and the vector
fields are neglected. In the spinodal region since it is expected to
have a small contribution, we neglect the frequency terms in the
propagators, i.e., $-\omega^{2}+k^{2}+m_{\textrm{s}}^{2}\approx
k^{2}+m_{\textrm{s}}^{2}$ and
$-\omega^{2}+k^{2}+m_{\textrm{\textrm{v}}}^{2}\approx
k^{2}+m_{\textrm{v}}^{2}$. Small fluctuations of the baryon density,
the scalar density and the current density are related to the
fluctuation of phase-space distribution function
$\delta\tilde{f}(\vec{k},\vec{p},\omega)$ according to,
\begin{eqnarray}\label{e14-e15}
    \delta\tilde{\rho}_{\textrm{b}}(\vec{k},\omega)&=&
    \gamma\int\frac{d^{3}p}{(2\pi)^{3}}\delta\tilde{f}(\vec{k},\vec{p},\omega), \\
\delta\tilde{\rho}_{\textrm{s}}(\vec{k},\omega)&=&
\gamma\int\frac{d^{3}p}{(2\pi)^{3}}
   \left[\delta\left(\frac{M^{*}}{e^{*}}\right)f_{0}(p)+
   \frac{M_{0}^{*}}{e_{0}^{*}}\delta\tilde{f}(\vec{k},\vec{p},\omega)\right] \nonumber \\
   &=& \gamma\int\frac{d^{3}p}{(2\pi)^{3}}\left[\left(\tilde{g}_{\textrm{v}}^{2}\frac{M_{0}^{*}}{e_{0}^{*3}}\vec{p}
   \cdot\delta\vec{\tilde{\rho}}_{\textrm{v}}(\vec{k},\omega)
   -\tilde{g}_{\textrm{s}}^{2}\frac{p^{2}}{e_{0}^{*3}}\delta\tilde{\rho}_{\textrm{s}}(\vec{k},\omega)\right)f_{0}(p)
   +\frac{M_{0}^{*}}{e_{0}^{*}}\delta\tilde{f}(\vec{k},\vec{p},\omega)\right]
   \nonumber \\
\end{eqnarray}
and
\begin{eqnarray}\label{e16}
   &&\delta\vec{\tilde{\rho}}_{\textrm{v}}(\vec{k},\omega)= \gamma\int\frac{d^{3}p}{(2\pi)^{3}}
   \left[\delta\left(\frac{\vec{p}^{*}}{e^{*}}\right)f_{0}(p)+
   \frac{\vec{p}}{e_{0}^{*}}\delta\tilde{f}(\vec{k},\vec{p},\omega)\right] \nonumber \\
   =&& \gamma\int\frac{d^{3}p}{(2\pi)^{3}}\left[\left(\tilde{g}_{\textrm{v}}^{2}\frac{\vec{p}}{e_{0}^{*3}}\vec{p}\cdot\delta\vec{\tilde{\rho}}_{\textrm{v}}(\vec{k},\omega)
   -\tilde{g}_{\textrm{v}}^{2}\frac{\delta\vec{\tilde{\rho}}_{\textrm{v}}(\vec{k},\omega)}{\epsilon_{0}^{*}}+
   \tilde{g}_{\textrm{s}}^{2}\frac{M_{0}^{*}}{e_{0}^{*3}}\vec{p}~\delta\tilde{\rho}_{\textrm{s}}(\vec{k},\omega)\right)f_{0}(p)
   +\frac{\vec{p}}{\epsilon_{0}^{*}}\delta\tilde{f}(\vec{k},\vec{p},\omega)\right].\nonumber  \\
\end{eqnarray}
Multiplying both sides of Eq. (\ref{e13}) by $M_{0}^{*}/e^{*}_{0}$,
1, $\vec{p}/e^{*}_{0}$ and integrating over the momentum, we deduce
a set of coupled algebraic equations for the small fluctuations of
the scalar density, the baryon density and the current density,
which can be put in to a matrix  form. Here we investigate spinodal
dynamics of the longitudinal unstable modes. For longitudinal modes
the current density oscillates along the direction of propagation,
$\delta\vec{\tilde{\rho}}_{\textrm{v}}(\vec{k},\omega)=\delta\tilde{\rho}_{\textrm{v}}(\vec{k},\omega)\hat{k}$.
Then, for the longitudinal modes, the set of equations become,
\begin{equation}\label{e17}
    \left(
      \begin{array}{ccc}
        A_{1} & A_{2} & A_{3} \\
        B_{1} & B_{2} & B_{3} \\
        C_{1} & C_{2} & C_{3} \\
      \end{array}
    \right)\left(
             \begin{array}{c}
               \delta\tilde{\rho}_{\textrm{v}}(\vec{k},\omega) \\
               \delta\tilde{\rho}_{\textrm{s}}(\vec{k},\omega) \\
               \delta\tilde{\rho}_{\textrm{b}}(\vec{k},\omega) \\
             \end{array}
           \right)=i\left(
             \begin{array}{c}
               \tilde{S}_{\textrm{b}}(\vec{k},\omega) \\
               \tilde{S}_{\textrm{s}}(\vec{k},\omega) \\
               \tilde{S}_{\textrm{v}}(\vec{k},\omega) \\
             \end{array}
           \right)
\end{equation}
where the element of the coefficient matrix are defined according
to,
\begin{equation}\label{e18}
    \left(
      \begin{array}{ccc}
        A_{1} & A_{2} & A_{3} \\
        B_{1} & B_{2} & B_{3} \\
        C_{1} & C_{2} & C_{3} \\
      \end{array}
    \right)=\left(
              \begin{array}{ccc}
                -\tilde{g}_{\textrm{v}}^{2}\chi_{\textrm{v}}(\vec{k},\omega) & -\tilde{g}_{\textrm{s}}^{2}\chi_{s}(\vec{k},\omega) & 1+\tilde{g}_{\textrm{v}}^{2}\chi_{\textrm{b}}(\vec{k},\omega) \\
                -\tilde{g}_{\textrm{v}}^{2}\tilde{\chi}_{\textrm{v}}(\vec{k},\omega) & 1+\tilde{g}_{\textrm{s}}^{2}\tilde{\chi}_{\textrm{s}}(\vec{k},\omega) & +\tilde{g}_{\textrm{v}}^{2}\chi_{\textrm{s}}(\vec{k},\omega) \\
                1+\tilde{g}_{\textrm{v}}^{2}\tilde{\chi}_{\textrm{b}}(\vec{k},\omega) & -\tilde{g}_{\textrm{s}}^{2}\chi_{\textrm{v}}(\vec{k},\omega) & +\tilde{g}_{\textrm{v}}^{2}\chi_{\textrm{v}}(\vec{k},\omega) \\
              \end{array}
            \right).
\end{equation}
In this expression, $\chi_{\textrm{b}}(\vec{k},\omega),
\chi_{\textrm{s}}(\vec{k},\omega)$ and
$\chi_{\textrm{v}}(\vec{k},\omega)$ denote the long wavelength limit
of relativistic Lindhard functions associated with baryon, scalar
and current density distribution functions,
\begin{equation}\label{e19}
    \left(
  \begin{array}{c}
    \chi_{\textrm{v}}(\vec{k},\omega) \\
    \chi_{\textrm{s}}(\vec{k},\omega) \\
    \chi_{\textrm{b}}(\vec{k},\omega) \\
  \end{array}
\right)=\gamma\int\frac{d^{3}p}{(2\pi\hbar)^{3}}\left(
  \begin{array}{c}
    \vec{p}\cdot \hat{k}/e_{0}^{*}\\
    M_{0}^{*}/e_{0}^{*} \\
    1 \\
  \end{array}
\right)\frac{\vec{k}\cdot\vec{\nabla}_{p}f_{0}(p)}{\omega-\vec{\textrm{v}}_{0}\cdot\vec{k}},
\end{equation}
and the stochastic source terms are determined by
\begin{equation}\label{e20}
    \left(
  \begin{array}{c}
    \tilde{S}_{\textrm{b}}(\vec{k},\omega) \\
    \tilde{S}_{\textrm{s}}(\vec{k},\omega) \\
    \tilde{S}_{\textrm{v}}(\vec{k},\omega) \\
  \end{array}
\right)=\gamma\int\frac{d^{3}p}{(2\pi)^{3}}\left(
  \begin{array}{c}
    1 \\
    M_{0}^{*}/e_{0}^{*} \\
    \vec{p}\cdot \hat{k}/e_{0}^{*} \\
  \end{array}
\right)\frac{\delta
\tilde{f}(\vec{k},\vec{p},0)}{\omega-\vec{\textrm{v}}_{0}\cdot\vec{k}}
\end{equation}
Other three elements of the coefficient matrix in Eq. (\ref{e18})
are given by,
\begin{equation}\label{e21}
    \tilde{\chi}_{\textrm{s}}(\vec{k},\omega)=\gamma\int\frac{d^{3}p}{(2\pi)^{3}}
    \left[\frac{p^{2}}{e_{0}^{*3}}f_{0}(p)-
    \frac{M_{0}^{*2}}{e_{0}^{*2}}
    \frac{\vec{k}\cdot\vec{\nabla}_{p}f_{0}(p)}{\omega-\vec{\textrm{v}}_{0}\cdot\vec{k}}\right],
\end{equation}
\begin{equation}\label{e22}
    \tilde{\chi}_{\textrm{v}}(\vec{k},\omega)=\gamma\int\frac{d^{3}p}{(2\pi)^{3}}\vec{p}\cdot\hat{k}
    \left[\frac{M_{0}^{*}}{e_{0}^{*2}}
    \frac{\vec{k}\cdot\vec{\nabla}_{p}f_{0}(p)}{\omega-\vec{\textrm{v}}_{0}\cdot\vec{k}}\right],
\end{equation}
and
\begin{equation}\label{e23}
    \tilde{\chi}_{\textrm{b}}(\vec{k},\omega)=\gamma\int\frac{d^{3}p}{(2\pi)^{3}}
    \left[\frac{e_{0}^{*2}-(\vec{p}\cdot\hat{k})^{2}}{e_{0}^{*3}}f_{0}(p)-
    \frac{(\vec{p}\cdot\hat{k})^{2}}{e_{0}^{*2}}
    \frac{\vec{k}\cdot\vec{\nabla}_{p}f_{0}(p)}{\omega-\vec{\textrm{v}}_{0}\cdot\vec{k}}\right].
\end{equation}
We obtain the solutions by inverting the algebraic matrix equation,
which gives for the baryon density fluctuations,
\begin{equation}\label{e24}
    \delta\tilde{\rho}_{B}(\vec{k},\omega)=i\frac{D_{1}\tilde{S}_{\textrm{b}}(\vec{k},\omega) + D_{2}\tilde{S}_{\textrm{s}}(\vec{k},\omega)
    + D_{3}\tilde{S}_{\textrm{v}}(\vec{k},\omega)}{\varepsilon(\vec{k},\omega)}
\end{equation}
where $D_{1}=B_{1}C_{2}-B_{2}C_{1}$, $D_{2}=C_{1}A_{2}-C_{2}A_{1}$
and $D_{3}=A_{1}B_{2}-A_{2}B_{1}$ and the quantity
$\varepsilon(\vec{k},\omega)=A_{3}D_{1}+B_{3}D_{2}+C_{3}D_{3}$
denotes the susceptibility.
\begin{figure}[h!]
       \includegraphics[width=10cm,height=12cm]{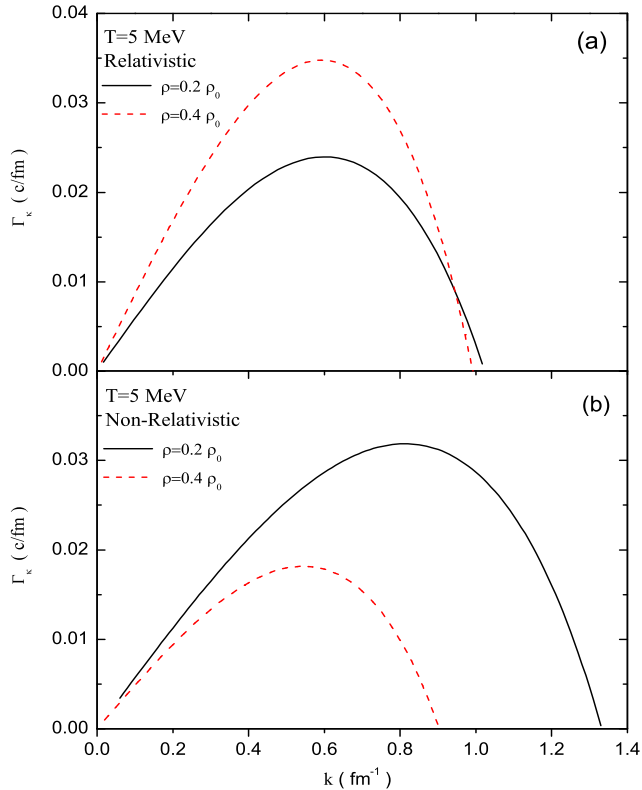}
       \caption{\label{fig1}Growth rates of unstable modes as a function of wave numbers in the spinodal region at baryon densities
    $\rho_{\textrm{b}}=0.2~\rho_{0}$ and $\rho_{\textrm{b}}=0.4 ~\rho_{0}$ at temperature $T=5 MeV$, (a) relativistic calculations, and (b) non-relativistic calculations.}
\end{figure}
\begin{figure}[h!]
      \includegraphics[width=10cm]{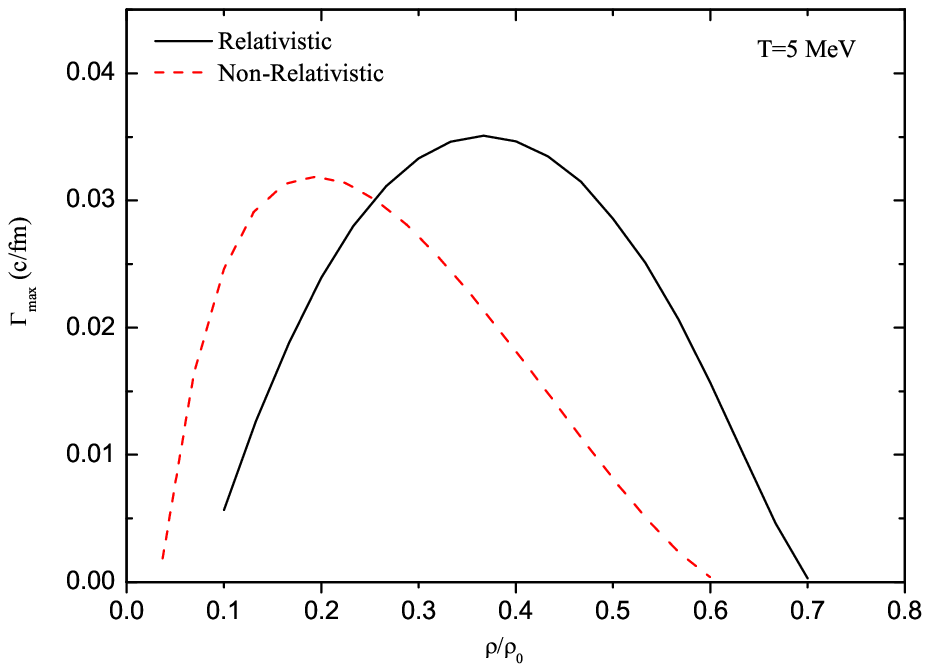}
       \caption{\label{fig2} Growth rates of the most unstable
modes as function of baryon density in spinodal region at
temperature $T=5 MeV$ in relativistic calculations (solid line) and
in non-relativistic calculations (dashed line).}
\end{figure}

\begin{figure}[h!]
     \includegraphics[width=10cm]{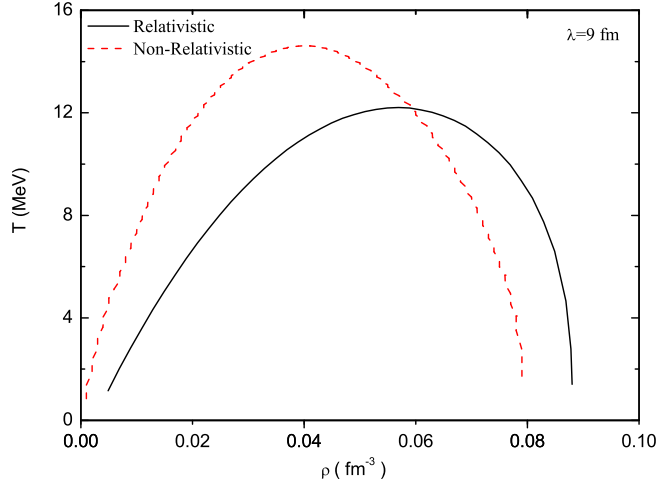}
    \caption{\label{fig3}Boundary of spinodal region in baryon density-temperature plane for the unstable mode
     with wavelengths $\lambda=9 ~ fm$  in relativistic calculations (solid line) and
in non-relativistic calculations (dashed line).}
\end{figure}

The evolution in time is determined by taking the inverse Fourier
transformation in time, which can be calculated with the help of
residue theorem \cite{R24}. Keeping only the growing and decaying
collective poles, we find,
\begin{equation}\label{e25}
    \delta\tilde{\rho}_{\textrm{b}}(\vec{k},t)=
    \delta\rho_{\textrm{b}}^{+}(\vec{k})e^{+\Gamma_{k}t}+\delta\rho_{\textrm{b}}^{-}(\vec{k})e^{-\Gamma_{k}t}
\end{equation}
Here, the amplitudes of baryon density fluctuations associated with
the growing and decaying modes at the initial instant are given by,
\begin{equation}\label{e26}
    \delta\rho_{\textrm{b}}^{\mp}(\vec{k}) = -\left\{\frac{D_{1}\tilde{S}_{\textrm{b}}(\vec{k},\omega)
  +D_{2}\tilde{S}_{\textrm{s}}(\vec{k},\omega)+D_{3}\tilde{S}_{\textrm{v}}(\vec{k},\omega)}{\partial\varepsilon(\vec{k},\omega)/\partial\omega}\right\}_{\omega=\mp i\Gamma_{k}}
\end{equation}
The growth and decay rates of the modes are obtained from the
dispersion relations, $\varepsilon(\vec{k},\omega)=0$, i.e. from the
roots of susceptibility. Solutions for the scalar density
fluctuations  $\delta\tilde{\rho}_{\textrm{s}}(\vec{k},\omega)$ and
the current density
$\delta\tilde{\rho}_{\textrm{v}}(\vec{k},\omega)$ fluctuations can
be expressed in a similar manner. In the original Walecka model,
there are four free parameters, coupling constants and meson masses.
The binding energy per nucleon at saturation density determines the
ratios of coupling constants to masses. The standard values of the
ratios $g_{\textrm{v}}^2(M/m_{\textrm{v}})^2=273.8$ and
$g_{\textrm{s}}^2(M/m_{\textrm{s}})^2=357.4$ give binding energy per
nucleon $15.75 ~MeV$ at saturation density \cite{R13,R14}. These
ratios lead to an effective nucleon mass $M_{0}^{*}=0.541 M$ and a
compressibility of $540~ MeV$ at the saturation density. In
numerical calculations, we take for the vector meson mass
$m_{\textrm{v}}=783 ~MeV$, and for the scalar meson mass,
$m_{\textrm{s}}=500 ~MeV$. As an example, the upper panel in Fig.
\ref{fig1} shows the growth rates of unstable modes as a function of
wave number in the spinodal region corresponding to the initial
baryon density $\rho_{\textrm{b}}=0.2 ~\rho_{0}$ and
$\rho_{\textrm{b}}=0.4 ~\rho_{0}$ at a temperature $T=5 ~MeV$. The
lower panel of Fig. \ref{fig1} illustrates the dispersion relations
obtained in the non-relativistic approach with an effective Skyrme
force \cite{R12}. Although direct comparison of these calculations
is rather difficult, we observe there are qualitative differences in
both calculations. The range of most unstable modes in relativistic
calculations is concentrated around $k=0.6 ~fm^{-1}$ in both
densities, while most unstable modes shift towards larger wave
numbers around $k=0.8 ~fm^{-1}$ at density $\rho_{\textrm{b}}=0.2
~\rho_{0}$ towards smaller wave numbers around $k=0.5 ~fm^{-1}$ at
density $\rho_{\textrm{b}}=0.4 ~\rho_{0}$. Growth rates of most
unstable modes at density $\rho_{\textrm{b}}=0.4 ~\rho_{0}$  in
relativistic calculations are nearly factor of two larger than those
results obtained in the non-relativistic calculations, while at low
density $\rho_{\textrm{b}}=0.2 ~\rho_{0}$ the growth rates are
smaller in relativistic calculations. Fig. \ref{fig2} illustrates
growth rates of the most unstable modes as a function of density in
both relativistic and non-relativistic approaches. We observe the
qualitative difference in the unstable response of the system: the
system exhibits most unstable behavior at higher densities around
$\rho_{\textrm{b}}=0.4 ~\rho_{0}$ in the relativistic approach while
most unstable behavior occurs in the non-relativistic calculations
at lower densities around $\rho_{\textrm{b}}=0.2 ~\rho_{0}$. As an
example of phase diagrams, Fig. \ref{fig3} shows the boundary of
spinodal region for the unstable mode of wavelength $\lambda=9.0
~fm$. Again, we observe that the unstable behavior shifts towards
higher densities in relativistic calculations.

\subsection{Growth of Density fluctuations}
In this section, we calculate the early growth of baryon density
fluctuations in nuclear matter. Spectral intensity of density
correlation function $\tilde{\sigma}_{\textrm{bb}}(\vec{k},t)$ is
related to the variance of Fourier transform of baryon density
fluctuation according to,
\begin{eqnarray}\label{e27}
  \tilde{\sigma}_{\textrm{bb}}(\vec{k},t)(2\pi)^{3} \delta(\vec{k}-\vec{k}')=
  \overline{\delta\tilde{\rho}_{\textrm{b}}(\vec{k},t)\delta\tilde{\rho}_{\textrm{b}}^{*} (\vec{k}',t)}
\end{eqnarray}
We calculate the spectral function using the solution (\ref{e25})
and the expression (\ref{e7}) for the initial fluctuations to give,
\begin{equation}\label{e28}
    \tilde{\sigma}_{\textrm{bb}}(\vec{k},t)=\frac{E_{\textrm{b}}^{+}(\vec{k})}{|[\partial\epsilon(\vec{k},\omega)/\partial\omega]_{\omega=i\Gamma_{k}}|^{2}}
    (e^{2\Gamma_{k} t}+e^{-2\Gamma_{k} t})+\frac{2E_{\textrm{b}}^{-}(\vec{k})}{|[\partial\epsilon(\vec{k},\omega)/\partial\omega]_{\omega=i\Gamma_{k}}|^{2}}
\end{equation}
where
\begin{eqnarray}\label{e29}
  E_{\textrm{b}}^{\mp}(\vec{k}) &=& |D_{1}|^{2}K_{11}^{\mp}+|D_{2}|^{2}K_{22}^{\mp}\mp|D_{3}|^{2}K_{33}^{\mp}+2D_{1}D_{2}K_{12}^{\mp}
\end{eqnarray}
with
\begin{equation}\label{e30}
    K_{11}^{\mp}=\gamma^{2}\int\frac{d^{3}p}{(2\pi)^{3}}\frac{\Gamma_{k}^{2}
    \mp(\vec{\textrm{v}}_0\cdot\vec{k})^{2}}{[\Gamma_{k}^{2}
    +(\vec{\textrm{v}}_0\cdot\vec{k})^{2}]^{2}}f_{0}(p)[1-f_{0}(p)],
\end{equation}
\begin{equation}\label{e31}
    K_{22}^{\mp}=\gamma^{2}\int\frac{d^{3}p}{(2\pi)^{3}}\left(\frac{M_{0}^{*}}{e^{*}_{0}}\right)^{2}\frac{\Gamma_{k}^{2}
    \mp(\vec{\textrm{v}}_0\cdot\vec{k})^{2}}{[\Gamma_{k}^{2}
    +(\vec{\textrm{v}}_0\cdot\vec{k})^{2}]^{2}}f_{0}(p)[1-f_{0}(p)],
\end{equation}
\begin{equation}\label{e32}
    K_{33}^{\mp}=\gamma^{2}\int\frac{d^{3}p}{(2\pi)^{3}}\left(\frac{\vec{p}\cdot\vec{k}}{e^{*}_{0}}\right)^{2}\frac{\Gamma_{k}^{2}
    \mp(\vec{\textrm{v}}_0\cdot\vec{k})^{2}}{[\Gamma_{k}^{2}
    +(\vec{\textrm{v}}_0\cdot\vec{k})^{2}]^{2}}f_{0}(p)[1-f_{0}(p)],
\end{equation}
and
\begin{equation}\label{e33}
    K_{12}^{\mp}=\gamma^{2}\int\frac{d^{3}p}{(2\pi)^{3}}\frac{M_{0}^{*}}{e^{*}_{0}}\frac{\Gamma_{k}^{2}
    \mp(\vec{\textrm{v}}_0\cdot\vec{k})^{2}}{[\Gamma_{k}^{2}
    +(\vec{\textrm{v}}_0\cdot\vec{k})^{2}]^{2}}f_{0}(p)[1-f_{0}(p)].
\end{equation}

\begin{figure}[b!]
      \includegraphics[width=10cm,height=12cm]{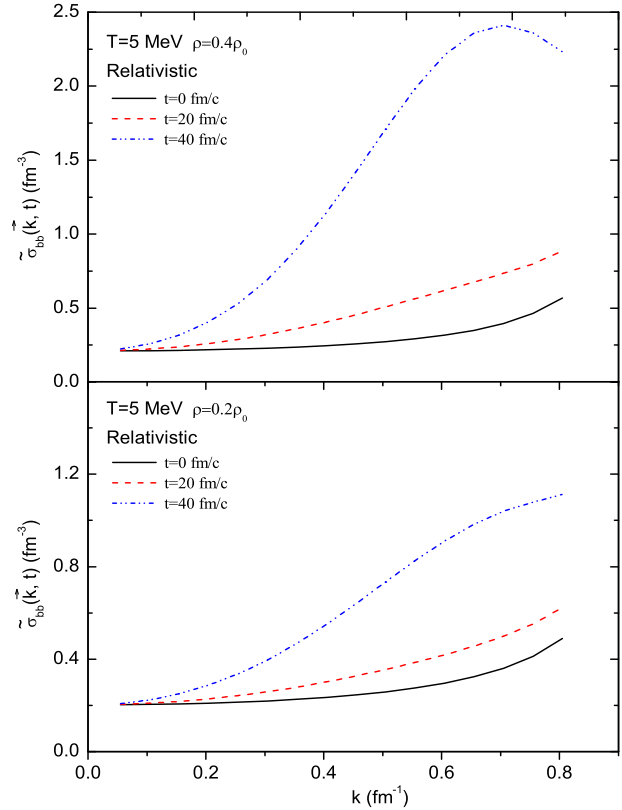}
       \caption{\label{fig4}Spectral intensity $\tilde{\sigma}_{\textrm{b}}(\vec{k},t)$ of baryon density correlation function as a function
    of wave number at times $t=0$, $t=20 ~ fm/c$ and
    $t=40 ~ fm/c $ at temperature $T=5 ~MeV$ in relativistic calculations
    at density (a) $\rho_{\textrm{b}}=0.2 ~\rho_{0}$ and (b) $\rho_{\textrm{b}}=0.4 ~\rho_{0}$.}
\end{figure}

\begin{figure}[h!]
      \includegraphics[width=10cm,height=12cm]{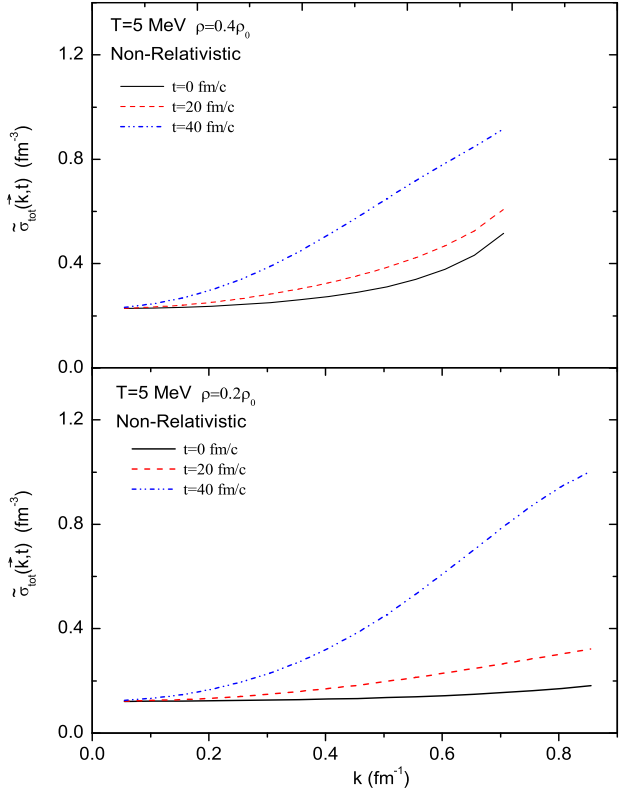}
       \caption{\label{fig5}Same as figure \ref{fig4} in non-relativistic calculations.}
\end{figure}
~~~~~~~\\
Upper and lower panels of Fig. \ref{fig4} show the spectral
intensity of the baryon density correlation function as a function
of wave number at times $t=0$, $t=20 ~ fm/c$ and $t=40 ~ fm/c$ at
temperature $T=5 ~MeV$ in relativistic calculations at densities
$\rho_{\textrm{b}}=0.2 ~\rho_{0}$ and $\rho_{\textrm{b}}=0.4
~\rho_{0}$, respectively. We observe that the largest growth occurs
over the range of wave numbers corresponding to the range of
dominant unstable modes. Spectral intensity in the vicinity of most
unstable modes of $k=0.6~fm^{-1}$ grows about a factor of ten at
density $\rho_{\textrm{b}}=0.2 ~\rho_{0}$ and about a factor of six
at density $\rho_{\textrm{b}}=0.4 ~\rho_{0}$ during the time
interval of $t=40 ~ fm/c$. Fig. \ref{fig5} shows the similar
information calculated in non-relativistic approaches. We notice
that at density $\rho_{\textrm{b}}=0.2 ~\rho_{0}$ the behavior of
spectral intensity is rather similar in relativistic and
non-relativistic approache. However, at higher density
$\rho_{\textrm{b}}=0.4 ~\rho_{0}$, the spectral intensity grows
slower in the non-relativistic calculations than those obtained in
the relativistic approach. We note that in determining time
evolution $\delta\rho_{\textrm{b}}(\vec{k},t)$ with the help of the
residue theorem, there are other contributions arising form the
non-collective pole of the susceptibility
$\varepsilon(\vec{k},\omega)$ and from the poles of source terms
$\tilde{S}_{\textrm{v}}(\vec{k},\omega)$,
$\tilde{S}_{\textrm{s}}(\vec{k},\omega)$ and
$\tilde{S}_{\textrm{b}}(\vec{k},\omega)$. These contributions, in
particular towards the short wavelengths, i.e. towards higher wave
numbers, are important at the initial stage, however they damp out
in a short time interval \cite{R25}. Since, we do not include
effects from non-collective poles, we terminate the spectral in Fig.
\ref{fig5} at a cut-off wave number $k_{c}~\approx~0.7~
fm^{-1}~-~0.8~ fm^{-1}$. Consequently, the expression (\ref{e28})
provides a good approximation for $\tilde{\sigma}_{bb}(\vec{k},t)$
in the long wavelength regime below $k_{c}$.

\begin{figure}[b!]
       \includegraphics[width=10cm,height=12cm]{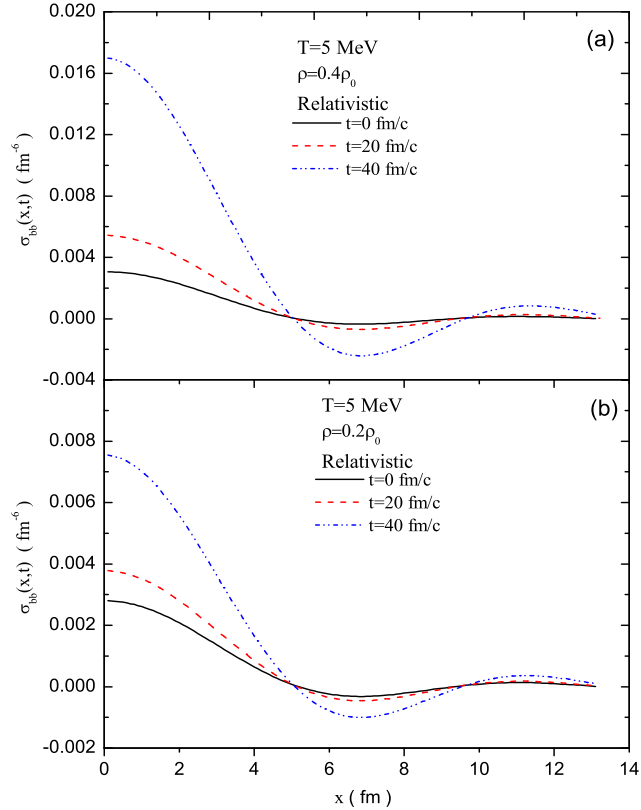}
      \caption{\label{fig6}Baryon density correlation function $\sigma_{\textrm{b}}(x,t)$ as a function of distance $x=|\vec{r}-\vec{r}'|$
    between two space points at times $t=0$, $t=20 ~ fm/c$ and $t=40 ~ fm/c$ at temperature $T= ~5$ MeV in
    relativistic calculations  at density (a) $\rho_{\textrm{b}}=0.2 ~\rho_{0}$ and
    (b) $\rho_{\textrm{b}}=0.4 ~\rho_{0}$.}
\end{figure}
\begin{figure}[h!]
      \includegraphics[width=10cm,height=12cm]{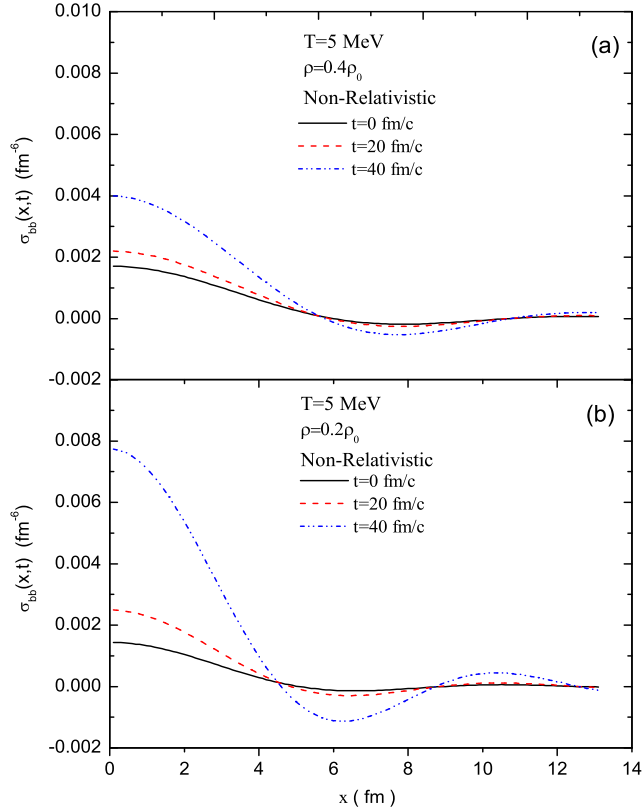}
       \caption{\label{fig7}Same as figure \ref{fig6} in non-relativistic calculations.}
\end{figure}

~~~~~~\\
Local baryon density fluctuations
$\delta\rho_{\textrm{b}}(\vec{r},t)$ are determined by the Fourier
transform of $\delta\rho_{\textrm{b}}(\vec{k},t)$. Equal time
correlation function of baryon density fluctuations as a function of
distance between two space locations can be expressed in terms of
the spectral intensity as,
\begin{equation}\label{e34}
    \sigma_{\textrm{bb}}(|\vec{r}-\vec{r}'|,t)=\overline{\delta n{\textrm{b}}(\vec{r},t)\delta n_{\textrm{b}}(\vec{r}',t)}=
    \int\frac{d^{3}k}{(2\pi)^{3}}e^{i\vec{k}\cdot\vec{r}}\tilde{\sigma}_{\textrm{bb}}(\vec{k},t).
\end{equation}
The baryon density correlation function carries useful information
about the unstable dynamics of the matter in the spinodal region. As
an example, the upper and lower panels of Fig. \ref{fig6}
illustrates the baryon density correlation function as a function
distance between two space points at times $t=0$, $t=20 ~ fm/c$, and
$t=40 ~ fm/c$ at temperature $T= ~ 5$ MeV in relativistic
calculations at densities $\rho_{\textrm{b}}=0.4 ~\rho_{0}$ and
$\rho_{\textrm{b}}=0.2 ~\rho_{0}$, respectively. Complementary to
the dispersion relation, correlation length of baryon density
fluctuations provides an additional measure for the size of the
primary fragmentation pattern. We can estimate the correlations
length of baryon density fluctuations as the width of the
correlation function at at half maximum. From the figure, we
estimate that the correlation length is about the same at both
densities and temperatures around $3.0 ~fm$, which is consistent
with the dispersion relation presented in Fig. \ref{fig1}. Baryon
density fluctuations grow faster at $\rho_{\textrm{b}}=0.4
~\rho_{0}$ than $\rho_{\textrm{b}}=0.2 ~\rho_{0}$. Fig. \ref{fig7}
shows the similar information calculated in the non-relativistic
approach \cite{R12}. The correlation length is around $3.0 ~fm$ at
$\rho_{\textrm{b}}=0.4 ~\rho_{0}$ and $3.0 ~fm$ at the lower density
$\rho_{\textrm{b}}=0.2 ~\rho_{0}$. However, unlike the relativistic
calculations, the baryon density fluctuations grow faster at lower
density $\rho_{\textrm{b}}=0.2 ~\rho_{0}$ than at
$\rho_{\textrm{b}}=0.4 ~\rho_{0}$, which is consistent result
presented in Fig. \ref{fig2}.

\section{CONCLUSIONS}
It has been demonstrated in recent publications \cite{R7,R9,R10,R12}
that the stochastic mean-field approach incorporates both the
one-body dissipation and the associated fluctuation mechanism in a
manner consistent with the fluctuation-dissipation theorem of
non-equilibrium statistical mechanics. Therefore the approach
provides a powerful tool for investigating dynamics of density
fluctuations in low-energy nuclear collisions. In a similar manner,
it is possible to develop an extension of the relativistic
mean-field theory by incorporating the initial quantal zero point
fluctuations and thermal fluctuations of density in a stochastic
manner. In this work, by employing the stochastic extension of the
relativistic mean-field approach, we investigate spinodal
instabilities in symmetric nuclear matter in the semi-classical
framework. We determine the growth rates of unstable collective
modes at different initial densities and temperatures. Stochastic
approach also allows us to calculate early development of baryon
density correlation functions in spinodal region, which provides
valuable complementary information about the emerging fragmentation
pattern of the system. We compare the results with those obtained in
non-relativistic calculations under similar conditions. Our
calculations indicate a qualitative difference in behavior in the
unstable response of the system. In the relativistic approach, the
system exhibits most unstable behavior at higher baryon densities
around $\rho_{\textrm{b}}=0.4 ~\rho_{0}$, while in the
non-relativistic calculations most unstable behavior occurs at lower
baryon densities around $\rho_{\textrm{b}}=0.2 ~\rho_{0}$.  In the
present exploratory work, we employ the original Walacka model
without self-interaction of scalar meson. The qualitative difference
in the unstable behavior may be partly due to the fact that the
original Walecka model leads to a relatively small value of nucleon
effective mass of $M^*=0.541M$ and a large nuclear compressibility
of $540~MeV$. On the other hand, the Skyrme interaction that we
employ in non-relativistic calculations gives rise to a
compressibility of $201~MeV$ \cite{R12}. It will be interesting to
carry out further investigations of spinodal dynamics in symmetric
and charge asymmetric nuclear matter by including self-interaction
of the scalar meson and also including the rho meson in the
calculations. Inclusion of the self-interaction of scalar meson
allows us to investigate spinodal dynamics over a wide range of
nuclear compressibility and nuclear effective mass. We also note by
working in the semi-classical framework, we neglect the quantum
statistical effects on the baryon density correlation function,
which become important at lower temperatures and also at lower
densities.
\begin{acknowledgments}
S.A. gratefully acknowledges TUBITAK for a partial support and METU
for warm hospitality extended to him during his visit. This work is
supported in part by the US DOE grant No. DE-FG05-89ER40530 and in
part by TUBITAK grant No. 107T691.
\end{acknowledgments}

\end{document}